\documentclass[aps,pra,twocolumn,superscriptaddress,showpacs]{revtex4-1}
\usepackage{graphicx,bm,color,float,textcomp,mathtools}
\newcommand\sho{$\rm SrHo_2O_4$}

\newcommand\sto{$\rm SrTb_2O_4$}
\newcommand\seo{$\rm SrEr_2O_4$}
\newcommand\syoE{$\rm SrY_2O_4$:Er$^{3+}$}
\newcommand\syoEhalf{$\rm SrY_2O_4$:Er$^{3+}$ (0.5 at.\%)}
\newcommand\sdo{$\rm SrDy_2O_4$}
\newcommand\sybo{$\rm SrYb_2O_4$}
\newcommand\syo{$\rm SrY_2O_4$}
\newcommand\sRo{Sr$R_2$O$_4$}
\newcommand\afm{antiferromagnetic}

\begin{document}
\title{Magnetic and spectral properties of multi-sublattice oxides $\mathbf{SrY_2O_4:Er^{3+}}$ and $\mathbf{SrEr_2O_4}$}
\date{\today}
\author{B.~Z.~Malkin}
\author{S.~I.~Nikitin}
\author{I.~E.~Mumdzhi}
\author{D.~G.~Zverev}
\author{R.~V.~Yusupov}
\author{I.~F.~Gilmutdinov}
\author{R.~G.~Batulin}
\author{B.~F.~Gabbasov}
\author{A.~G.~Kiiamov}
\affiliation{Kazan Federal University, Kazan, 420008, Kremlevskaya 18, Russia}
\author{D.~T.~Adroja}
\affiliation{ISIS Facility, STFC Rutherford Appleton Laboratory, Chilton, Didcot, OX11 0QX, United Kingdom}
\affiliation{Highly Correlated Electron Group, Physics Department, University of Johannesburg, P.O. Box 524, Auckland Park 2006, South Africa}
\author{O.~Young}
\affiliation{High Field Magnet Laboratory, Institute for Molecules and Materials, Radboud University, 6525 ED Nijmegen, The Netherlands}
\affiliation{Department of Physics, University of Warwick, Coventry CV4 7AL, United Kingdom}
\author{O.~A.~Petrenko}
\email{O.Petrenko@warwick.ac.uk}
\affiliation{Department of Physics, University of Warwick, Coventry CV4 7AL, United Kingdom}
\begin{abstract} 
\seo\ is a geometrically frustrated magnet which demonstrates rather unusual properties at low temperatures including a coexistence of long- and short-range magnetic order, characterized by two different propagation vectors.
In the present work, the effects of crystal fields (CF) in this compound containing four magnetically inequivalent erbium sublattices are investigated experimentally and theoretically.
We combine the measurements of the CF levels of the Er$^{3+}$ ions made on a powder sample of \seo\ using neutron spectroscopy with site-selective optical and electron paramagnetic resonance measurements performed on single crystal samples of the lightly Er-doped nonmagnetic analogue, \syo.
Two sets of CF parameters corresponding to the Er$^{3+}$ ions at the crystallographically inequivalent lattice sites are derived which fit all the available experimental data well, including the magnetization and dc susceptibility data for both lightly doped and concentrated samples.       
\end{abstract}
\maketitle
\section{Introduction}
The family of rare-earth (RE) strontium oxides, \sRo,  has recently been identified as a useful addition to a small group of compounds where the combination of geometrical frustration, magnetic low-dimensionality and single-ion physics results in the stabilisation of highly unusual ground states at temperatures much lower than those expected from the strength of the magnetic interactions~\cite{Karunadasa_2005,Petrenko_2014}.
A wide variety of unconventional magnetic properties observed in these compounds include a coexistance of long-range \afm\ and short-range incommensurate order in \seo~\cite{Petrenko_2008,Hayes_2011}, two types of short-range order in \sho~\cite{Young_2012,Young_2013,Wen_2015}, noncollinear order in \sybo~\cite{Quintero_2012}, incommensurate magnetic structure in \sto~\cite{Li_2014_STO} and the absence of the longer-range magnetic correlations in \sdo\ down to the lowest experimentally available temperatures~\cite{Cheffings_2013}. 
To elucidate and model specific magnetic properties of these oxides, which can be described as geometrically frustrated multi-sublattice magnets with a quasi-one-dimensional space structure, it is necessary to establish the spectral properties of the magnetic subsystem formed by the RE ions.
However, only sparse information on low-energy electronic excitations in \sRo\ compounds has been obtained from studies of inelastic neutron scattering in \sho~\cite{Ghosh_2011,Poole_2014} and \sdo~\cite{Poole_2014}.
The present work is aimed at the determination of sets of parameters used in the microscopic models of spectral and magnetic properties of a single RE ion in a dielectric crystal.
With these parameters, it is possible to carry out a physically justified analysis of the different quantum effects caused by the interactions between the paramagnetic RE ions.
To characterize the single ion properties, we start from experimental and theoretical studies of a strongly diluted RE subsystem in the isostructural diamagnetic compound, \syo.
Studies of \syo\ single-crystals doped with the Er$^{3+}$ ions are well suited for this purpose due to the rich spectrum of electronic excitations of the Er$^{3+}$ ion lying in the spectral range accessible to site-selective optical measurements.

We have performed a series of spectroscopic studies of the Er$^{3+}$ ions in the \syoEhalf\ sample (measurements of the electron paramagnetic resonance, EPR, and optical spectra) and in the concentrated \seo\ crystals (measurements of the inelastic neutron scattering at different temperatures) as well as the magnetometry studies (measurements of the magnetization versus external magnetic fields applied along the principal crystallographic axes at different temperatures).
Analysis of the results of these measurements in the framework of a semi-phenomenological crystal field (CF) theory~\cite{Malkin_1987} and the derived four-particle self-consistent model of magnetic interactions have allowed us to determine the sets of CF parameters and exchange coupling constants related to the specific positions of the Er$^{3+}$ ions in the crystal lattice.
From comparisons of the simulated spectral envelopes of the inelastic neutron scattering, the temperature dependencies of the dc susceptibility tensor components and the field dependencies of the magnetization in \seo\ with the experimental data, it is evident that the spectral parameters of the impurity Er$^{3+}$ ions in \syo\ can be used successfully to reproduce spectral and magnetic properties of the erbium subsystem in \seo.

The paper is organized as follows: the experimental methods employed in the present work and the crystal structure of the samples are described in Sec.~\ref{sec_methods}.
The results of measurements are presented in Sec.~\ref{sec_results}.
Sec.~\ref{sec_modeling} contains an analysis of the data obtained which involves simulations of the crystal fields affecting the Er$^{3+}$ ions at different crystallographic sites and modeling of the magnetic properties of the concentrated \seo\ crystals in the paramagnetic phase.
The paper ends with a brief summary of the results.

\section{Experimental methods and crystal structure}
\label{sec_methods}
A single crystal of \syo\ doped with Er$^{3+}$ ions (0.5 at.\%) was grown by the optical floating zone technique.
SrCO$_{3}$ (Alfa Aesar, 99.99\%), Y$_2$O$_3$ (Alfa Aesar, 99.99\%) and Er$_2$O$_3$ (Alfa Aesar, 99.9\%) oxides were used as starting materials.
The thoroughly ground and mixed stoichiometric composition was initially sintered at 1050~$^\circ$C for 8~hours.
The synthesized powder was examined with powder X-ray diffraction (Bruker D8 ADVANCE, Cu K$\alpha$) and found to be a phase-pure material with lattice parameters and space group symmetry corresponding to the structure of \syo.
A cylindrical rod was formed from the powder using a hydrostatic press, and this rod was used as the ingot for the crystal growth.
The single crystal was grown in an air flow of 0.5~L/min at ambient pressure using an optical floating zone furnace FZ-T-4000-H-VII-VPO-PC (Crystal Systems Corp., Japan) equipped with four 1~kW halogen lamps at a rate of 3~mm/h.
The feed and seed rods were counterrotated at a rate of 15~rpm in order to obtain a homogeneous molten zone.
The grown crystal was pinkish in color and was of high optical quality.

\begin{figure}[tb] 
\centering
\includegraphics[width=0.88\columnwidth]{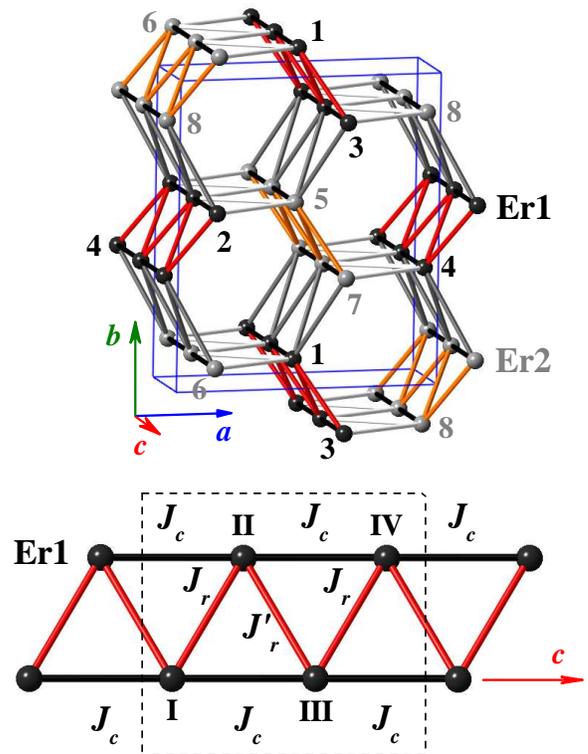}
\caption{(Color online) Top panel: Positions of the Er$^{3+}$ ions within \seo, with the two crystallographically inequivalent Er$^{3+}$ sites shown in different colors.
The blue box represents a crystal unit cell of the $Pnam$ space group.
Bottom panel: a schematic of the magnetic interactions between the Er$^{3+}$ ions in one of the zig-zag chains.
A four-particle cluster used to model the magnetization process is highlighted by the dashed box.}
\label{Fig1_structure}
\end{figure}

The crystal structure of \sRo\ is orthorhombic with the space group $Pnam$.
The unit cell has 28 ions in total, split into seven groups each containing four crystallographically equivalent $\lambda$-ions ($\lambda =$~Sr, R1, R2, O1, O2, O3 and O4) at the Wyckoff 4c positions.
The ions coordinates (in units of lattice constants $a$, $b$, and $c$),
$\bm{r}_{1,\lambda} = (x_{\lambda }, y_{\lambda }, 0.25)$,
$\bm{r}_{2,\lambda} = (0.5-x_\lambda,y_\lambda - 0.5,0.75)$,
$\bm{r}_{3,\lambda} = (-x_\lambda,-y_\lambda,0.75)$,
$\bm{r}_{4,\lambda} = (x_\lambda -0.5, 0.5-y_{\lambda },0.25)$ are determined by structure parameters $x_\lambda$ and $y_\lambda$.
The lattice constants and the structure parameters of \syo\ and \seo\ were reported in Refs.~\cite{Muller-Buschbaum_1968,Karunadasa_2005,Li_2014_SEO}.
In particular, $a=10.0166$~\AA, $b=11.8541$~\AA, and $c=3.3848$~\AA, $x_{\rm R1}=0.4227$, $y_{\rm R1}=0.1101$, and $x_{\rm R2}=0.4232$, $y_{\rm R2}=0.6118$ in \seo\ at~80~K~\cite{Li_2014_SEO}.
The impurity Er$^{3+}$ ions in \syo\ substitute for Y$^{3+}$ ions at the R1 and R2 sites and, as in \seo, have six nearest neighbor oxygen ions which form distorted octahedra.
The average distances between the Er$^{3+}$ and O$^{2-}$ ions in Er1O$_6$ and Er2O$_6$ octahedra in \seo\ are equal to 2.26 and 2.28~\AA, respectively, at 80~K.
Crystal fields at the R1 and R2 sites have point symmetry $C_s$.
The R1- and R2-sublattices are labeled below by the indices $\rho$ from 1 to 4 ($\bm{r}_\rho =\bm{r}_{\rho ,{\rm R1}}$) and from 5 to 8 ($\bm{r}_\rho = \bm{r}_{\rho -4,{\rm R2}}$), respectively (see Fig.~\ref{Fig1_structure}).
Ions with the indices $\rho =1$ and $\rho =3 $ (2 and 4, 5 and 7, 6 and 8) are connected by the inversion operation and form a ladder with legs parallel to the $c$ axis and shifted relative to one another by $c$/2.
In general, a ladder can be considered as a zigzag chain where distances between the neighboring ions lying on the adjacent legs, $d_1$, and on the same leg, $d_2$, are slightly different, and $d_1>d_2$.
Each leg in a ladder has two neighboring legs belonging to other ladders, and the projection of the R1 and R2 chains results in a honeycomb network (of edge-sharing distorted hexagons) in the $ab$~plane, see Fig.~\ref{Fig1_structure}.

The \syoEhalf\ sample used in the EPR, optical and magnetization studies was aligned using X-ray diffraction.
It had the shape of a rectangular parallelepiped with the dimensions of $2.5 \times 2 \times 1$~mm$^3$ with the edges along the crystallographic $a$, $b$ and $c$ axes, respectively.
EPR spectra were measured with the X-band Bruker ESP300 spectrometer.
The standard ER4102ST rectangular cavity with the TE$_{102}$ mode was used.
The studies were performed at 15~K, and the sample temperature was controlled using an Oxford Instruments ESR9 liquid helium flow system.

The magnetization measurements of the aligned \syoEhalf\ single crystal were performed using a Quantum Design Physical Property Measurement System (PPMS-9) Vibrating Sample Magnetometer (VSM).

Fluorescence of the crystals was excited with either a pulsed tunable dye laser (Littrow type oscillator and amplifier, linewidth of about 1 \AA ) pumped by the second or third harmonic of a Nd-YAG laser (LQ129, Solar LS) in
the visible spectral range or a Ti:Sapphire tunable laser with linewidth of about 0.4 \AA\ (LX325, Solar LS) pumped by the second harmonic of a Nd-YAG laser (LQ829, Solar LS) in the near-IR range.
The spectra were analyzed with an MDR-23 monochromator.
The fluorescence signal was detected by a cooled photomultiplier (PMT-106 or PMT-83) in the photon-counting regime. 
The studied \syoEhalf\ crystal was kept in helium vapor at a temperature of 4.2~K.

The polycrystalline sample of \seo\ used in the neutron scattering experiments was prepared from high-purity starting materials SrCO$_3$ and Er$_2$O$_3$ following Ref.~\cite{Karunadasa_2005}.
The same sample was previously used for high-resolution powder neutron diffraction experiments~\cite{Petrenko_2008} which have revealed the absence of any significant amount of impurities or chemical disorder in the samples.
The inelastic neutron scattering (INS) experiment was performed with the HET time-of-flight spectrometer at ISIS, STFC Rutherford Appleton Laboratory, United Kingdom in the temperature range 5 to 140~K.
For the INS experiment the powdered samples were wrapped in a thin Al-foil and mounted inside a thin-walled cylindrical Al-can.
The sample mount was cooled in a top-loading closed-cycle refrigerator with He-exchange gas and the INS data were collected using neutrons with incident energies $E_i$ between 24 and 120~meV, and for scattering angles between 3 and 135 degrees.
The elastic resolution (full width half maximum) was 1.5, 2.0 and 5.5~meV for the $E_i$ of 24, 40 and 120~meV, respectively.

Although magnetization data for \seo\ have been previously reported~\cite{Petrenko_2008,Hayes_2012} the emphasis for the earlier studies was either on the magnetization behavior in high magnetic field~\cite{Petrenko_2008} or on very low temperatures~\cite{Hayes_2012} where unusual plateaux have been observed rather than on the measurements of the absolute value of the magnetization.
We therefore took advantage of the much higher accuracy available with a SQUID magnetometer (MPMS, Quantum Design) compared to the VSM and extraction magnetometers previously used~\cite{Petrenko_2008} and remeasured magnetization in \seo\ at $T=5$~K for the field parallel to the three main crystallographical directions.
The sample preparation is described in Ref.~\cite{Balakrishnan_2009}.

\section{Experimental results}
\label{sec_results}
\subsection{EPR spectra}
\begin{figure}[tb] 
\centering
\includegraphics[width=0.88\columnwidth]{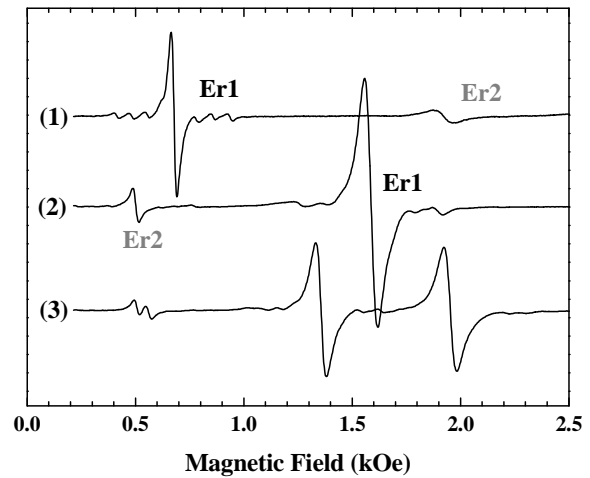}
\caption{EPR spectra ($T=15$~K, $\nu = 9.42265$~GHz) of the \syoE\ crystal for the magnetic field applied (1) along the $c$~axis, (2) along the $a$~axis, and (3) tilted by 20~degrees away from the $a$~axis in the $ab$~plane.}
\label{Fig2_EPR}
\end{figure}

Examples of the observed EPR spectra taken in crossed static ($\bm H$) and linearly polarized microwave ($\bm{H}_t$) magnetic fields, in particular, for static fields applied along the crystallographic $a$ and $c$ axis, are shown in Fig.~\ref{Fig2_EPR}.
For $\bm H \parallel c$ and $\bm{H}_t \parallel b$, two signals are detected in the spectrum at $H \sim 680$~Oe and $\sim 1920$~Oe.
The first signal marked as Er1 consists of the intense line originating from erbium isotopes with even mass numbers and a pronounced hyperfine octet due to the single isotope with an odd mass number, $^{167}$Er, which has a nuclear spin $I=7/2$ and the natural abundance of 22.9\%.
The second signal at 1920~Oe marked as Er2 is considerably broader than the first one.
As a consequence, it reveals a significantly less pronounced but still detectable hyperfine structure of $^{167}$Er$^{3+}$.
It should be emphasised that we cannot {\it a priori} identify the observed EPR signals with the R1 and R2 positions of the Er$^{3+}$ ions.
The identification of the EPR signals (as well as of the transitions in the optical spectra discussed below) was achieved on the basis of the crystal field calculations (see Sec.~\ref{sec_modeling}).
The integral intensity of the resonance absorption is proportional to the square of the $g$-tensor component corresponding to the microwave field direction.
The relationships between the intensities of the Er1 and Er2 EPR signals  (which are proportional to field derivatives of the absorption) in Fig.~\ref{Fig2_EPR} agree with the ratios of the corresponding $g$-factors of the erbium ions at the R1 and R2 sites determined below and presented in Table~\ref{TableI}.

With the magnetic field $\bm H$ applied along any of the crystallographic axes as well as lying in either the $ac$ or $bc$~plane, two signals are observed in the EPR spectra in agreement with the presence of the two crystallographically inequivalent R1 and R2 sites.
In an arbitrary oriented magnetic field tilted away from the $ac$ (or $bc$) plane, two magnetically inequivalent centers are revealed in the spectra for both the Er1 and Er2 positions (see Fig.~\ref{Fig2_EPR}, spectrum number 3).
\begin{table}[tb] 
\caption{$g$-factors of the ground doublets of the impurity Er$^{3+}$ ions in \syo\ (the orbital reduction factors used in the calculations are $k=0.985$ for Er1 and $k=0.97$ for Er2).}
\begin{center}
\begin{tabular}{|l|c|c|c|c|}
\hline
\hline
			& \multicolumn{2}{|c|}{ Er1 }	& \multicolumn{2}{|c|}{ Er2 }	\\ \cline{2-5}
			& Measured				& Calculated	& Measured			& Calculated	\\ \hline
$g_1$		& 5.382 $\pm$ 0.011			& 5.396		& 13.2 $\pm$ 0.1		& 13.22		\\
$g_2$		& 3.009 $\pm$ 0.007			& 3.009		& 1.69 $\pm$ 0.05		& 1.726		\\
$g_3=g_{cc}$	& 9.93 $\pm$ 0.03			& 9.944		& 3.42 $\pm$ 0.02		& 3.51		\\
$\phi$		& $(48.0 \pm 0.2)^\circ$		& $45.3^\circ$	& $(8.1 \pm 0.05)^\circ$	& $9.3^\circ$	\\
$g_{aa}$		& 4.239					& 4.357		& 13.07				& 13.05		\\
$g_{bb}$		& 4.478					& 4.38		& 2.502				& 2.73		\\ \hline
\end{tabular}
\end{center}
\label{TableI}
\end{table}

In a low-symmetry crystal field, all the energy levels of an Er$^{3+}$ ion with the ground electronic configuration $4f^{11}$ are Kramers doublets.
In an external magnetic field, the effective spin-Hamiltonian ($S=1/2$) of a Kramers doublet in the Cartesian system of coordinates with the coordinate axes along the principal axes of a $g$-tensor has the following form
\begin{equation}
\mathcal{H}_S = \mu_{\mathrm B} (g_1 H_1 \hat{S_1} + g_2 H_2 \hat{S_2} + g_3 H_3 \hat{S_3}).
\label{eq1_Ham}
\end{equation}
Here $\mu_{\mathrm B}$ is the Bohr magneton, $H_\alpha$ and $S_\alpha$ ($\alpha = 1, 2, 3$) are projections of the magnetic field and effective electronic spin moment on the coordinate axes, respectively.
In the case of a local $C_s$ symmetry of a paramagnetic ion, two principal axes (labeled by the indices 1 and 2) of the $g$-tensor are in the $ab$ (mirror) plane, and the third axis is parallel to the $c$ axis. 

In the EPR spectra with the frequency $\nu$ and magnetic field lying in the $ab$~plane at the angle $\vartheta$ to the $a$~axis, the resonant fields $H$ of erbium isotopes with even mass numbers are given by the equation
\begin{equation}
\mu_{\mathrm B} H = \frac{2 \sqrt{2} \pi \hbar \nu}{\sqrt{g^2_{1,\kappa } + g^2_{2,\kappa} + (g^2_{1,\kappa } - g^2_{2,\kappa}) \cos (2 \vartheta \pm 2 \varphi _{\kappa})}},
\label{eq2_field}
\end{equation}
where $+\varphi _\kappa$ and $-\varphi _\kappa$ are the angles between the principal axes corresponding to the $g$-factors $g_{1,\kappa}$ and the $a$~axis for the Er$^{3+}$ions at sites $\bm{r}_{1,\kappa}$, $\bm{r}_{3,\kappa}$ and $\bm{r}_{2,\kappa}$, $\bm{r}_{4,\kappa}$, respectively ($\kappa =$~R1, R2).
In agreement with the results of the measurements for an arbitrary orientation of the field $\bm H$, four EPR signals can be observed, but when the magnetic field is applied along a principal crystallographic axis $\alpha$ ($\alpha = a, b$ or $c$), there are only two lines corresponding to the resonant fields $H=2\pi \hbar \nu /(g_{\alpha \alpha , \kappa }\mu_{\mathrm B})$, where
\begin{eqnarray}
g_{aa,\kappa} & = & [g^2_{1,\kappa} + g^2_{2,\kappa} + (g^2_{1,\kappa} - g^2_{2,\kappa}) \cos (2\varphi _\kappa )]^{1/2}/\sqrt{2}, \hspace{5mm} \\
g_{bb,\kappa} & = & [g^2_{1,\kappa} + g^2_{2,\kappa} -  (g^2_{1,\kappa} - g^2_{2,\kappa}) \cos (2\varphi _\kappa )]^{1/2}/\sqrt{2}, \\ 
g_{cc,\kappa} & = & g_{3,\kappa}.
\label{eq35_g}
\end{eqnarray}
For the magnetic field in the $ac$~plane, the resonant fields satisfy the condition
\begin{equation}
\mu_{\mathrm B} H = 2 \pi \hbar \nu \lbrack g^2_{cc,\kappa } + (g^2_{aa,\kappa} - g^2_{cc,\kappa})\cos^2 \vartheta ]^{-1/2},
\label{eq6_field}
\end{equation}
where $\vartheta $ is the angle between the field and the $a$~axis.

\begin{figure}[tb] 
\centering
\includegraphics[width=0.88\columnwidth]{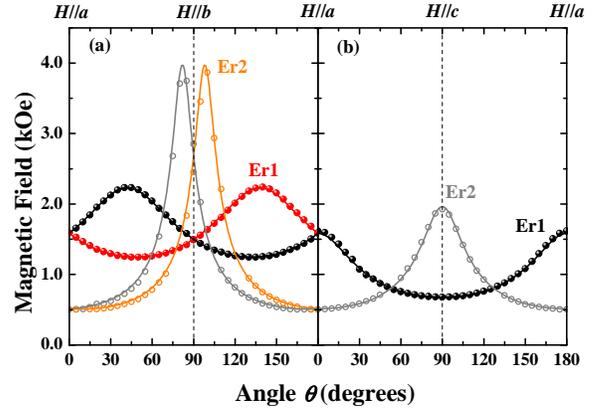}
\caption{(Color online) Measured (symbols) and calculated (lines) angular dependencies of the resonant magnetic field in the (a) $ab$ and (b) $ac$ planes of \syoEhalf\ at the frequency of 9.42265~GHz.
Black/red solid symbols and gray/orange open symbols are used for the Er1 and Er2 centers respectively.}
\label{Fig3_EPR}
\end{figure}
Angular dependencies of the EPR spectra were measured for the magnetic field lying in the $ab$ and $ac$~planes.
The resonant field dependence of the field direction for the erbium isotopes with even mass numbers are shown in Fig.~\ref{Fig3_EPR}.
These dependencies are well described by equations~(\ref{eq2_field}) and (\ref{eq6_field}).
The results of the fits to the experimental data, within the model of the isolated effective spin-1/2 doublet with the anisotropic $g$-tensor, are represented in Fig.~\ref{Fig3_EPR} by the solid lines.
The values of the principal $g$-tensor components, $g$-factors in the directions of the crystallographic axes as well as the angles between the principal directions of the $g$-tensor and the crystallographic axes are presented in Table~\ref{TableI}.
It should be noted that a possible misalignment of the sample in an external magnetic field may introduce errors in the obtained values of the $g$-factors of up to 0.1.
Nevertheless, the EPR studies reported here provide conclusive evidence for the strong magnetic anisotropy of the Er$^{3+}$ ions of an easy-axis type with the easy-axis along the $c$~direction at the R1 sites and along the two non-collinear directions in the $ab$~plane at the magnetically inequivalent R2 sites.

The available experimental data do not allow for a full characterization of the hyperfine interaction for any $^{167}$Er$^{3+}$ center.
The values of the hyperfine constants can be given with a satisfactory accuracy for only a limited range of the magnetic field directions with respect to the crystal axes.
The restrictions arise due to the strong broadening of the EPR lines.

Additional weak EPR signals were observed in the magnetic fields lying in the $ac$~plane.
The nature of the corresponding paramagnetic centers is not entirely clear, but it is possible that the impurity Er$^{3+}$ions substitute not only for Y$^{3+}$ but also for the Sr$^{2+}$ ions.

\subsection{Optical site-selective excitation and emission spectra}

\begin{figure}[tb] 
\centering
\includegraphics[width=0.88\columnwidth]{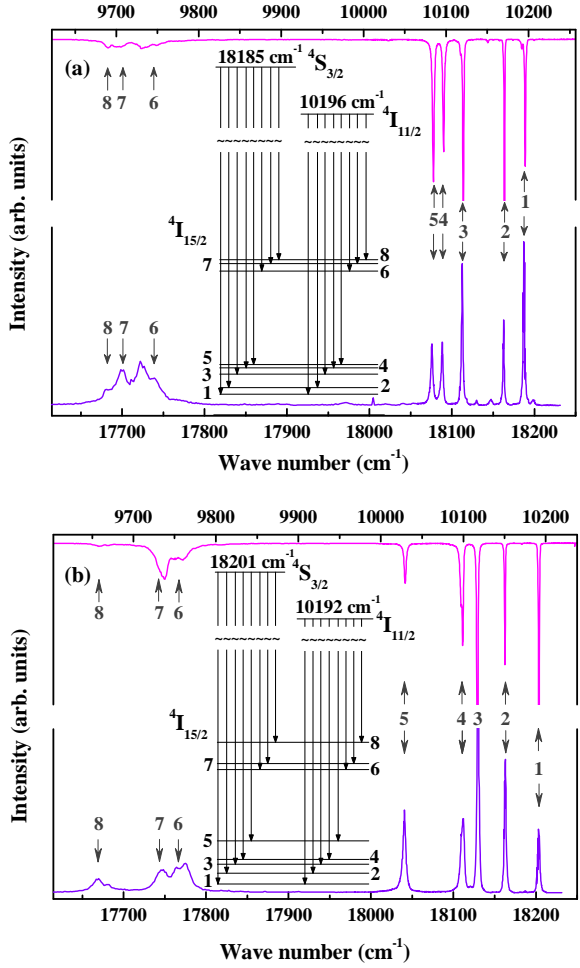}
\caption{(Color online) (a) Luminescence spectra at  $T=4.2$~K of the R1 centers corresponding to the emission from the lowest sublevels of $\rm ^4S_{3/2}$ (lower spectrum) and $\rm ^4I_{11/2}$ (upper spectrum) multiplets into the sublevels of the ground $\rm ^4I_{15/2}$ multiplet induced by radiation with the wave numbers of 19193 and 10355.5~cm$^{-1}$ absorbed due to the excitations of the Er$^{3+}$ ions into the sublevel number~4 in the $\rm ^2H_{11/2}$ multiplet or the sublevel number~6 in the $\rm ^4I_{11/2}$ multiplet, respectively (see Table~\ref{TableII}).
(b) Luminescence spectra of the R2 centers induced by radiation with the wave numbers of 19273.6 and 10393.9~cm$^{-1}$ absorbed due to the excitations into the sublevels number~6 in the $\rm ^2H_{11/2}$ and $\rm ^4I_{11/2}$ multiplets, respectively.
The insets give graphical representations of the crystal field energy levels for the two crystallographic sites.}
\label{Fig4_optics}
\end{figure}

We studied the emission of the \syoE\ samples for two channels corresponding to the radiative transitions from the two metastable states of the Er$^{3+}$ ions, namely, from the lowest CF sublevels of the $\rm ^4I_{11/2}$ and $\rm ^4S_{3/2}$ multiplets, to the sublevels of lower-lying multiplets.

The site-selective technique revealed the presence of two different sets of photoluminescent spectral lines originating from transitions in the Er$^{3+}$ ions (see Fig.~\ref{Fig4_optics}).
Here, energies of the CF sublevels of the ground ($\rm ^4I_{15/2}$) and the first excited ($\rm ^4I_{13/2}$) multiplets were determined from the selectively excited luminescence spectra.
The selective excitation of different centres is confirmed, in particular, by the very different luminescence spectra shown in Figs.~\ref{Fig4_optics}a and~\ref{Fig4_optics}b.

\begin{figure}[tb] 
\centering
\includegraphics[width=0.88\columnwidth]{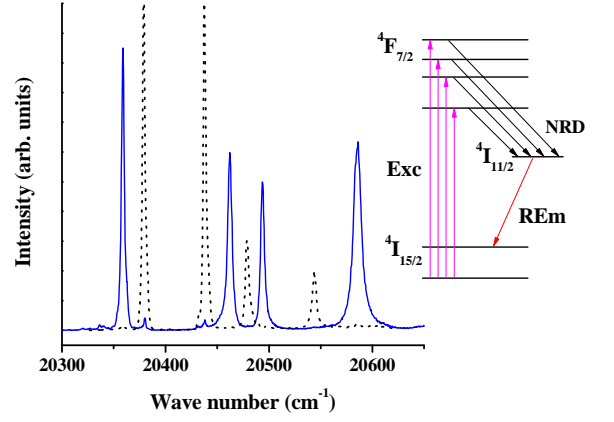}
\caption{(Color online) Excitation spectra of the R1 (solid line) and R2 (dashed line) centers in \syoEhalf\ corresponding to the transitions from the ground state into the $\rm ^4F_{7/2}$ multiplet of the Er$^{3+}$ ions at $T=4.2$~K registered by the emission from the lowest sublevels of the $\rm ^4I_{11/2}$ multiplet to the first excited sublevels of the ground multiplet at the wave numbers of 10170.7~cm$^{-1}$ (10195.9~cm$^{-1}$ -- 25.2 cm$^{-1}$) and 10150 cm$^{-1}$ (10191.1 cm$^{-1}$ -- 41.1 cm$^{-1}$), respectively.
The arrows in the inset show the excitations (Exc), the registered emission (REm) and the nonradiative stepped decay of the excited states (NRD).}
\label{Fig5_spectra}
\end{figure}
Additional spectral lines were observed with an intensity that was at least two orders of magnitude weaker.
These lines probably correspond to the Er$^{3+}$ions substituted for the Sr$^{2+}$ ions in the \syo\ lattice, however, the concentration of these centers is very small.
Energies of all CF sublevels of the $\rm ^4I_{11/2}$, $\rm ^4I_{9/2}$, $\rm ^4F_{9/2}$, $\rm ^2H_{11/2}$, $\rm ^4F_{7/2}$, $\rm ^4F_{5/2}$, $\rm ^4F_{3/2}$ and $\rm ^4S_{3/2}$ multiplets of both the R1 and R2 optical centers were determined from the excitation spectra of the luminescence measurements (an example is shown in Fig.~\ref{Fig5_spectra}).
The measured spectra provide evidence of a relatively weak exchange of the excitation energy between the R1 and R2 centers.

\begin{table}[H] 
\caption{Measured and calculated crystal field energies (in~cm$^{-1}$) of the Er$^{3+}$ ions in \syoEhalf\ and \seo\ (in brackets).}
\begin{center}
\begin{tabular}{|r|l|l|l|l|}
\hline
Multiplet			& \multicolumn{2}{|c|}{Er1}	& \multicolumn{2}{|c|}{Er2}	\\ \cline{2-5}
				& Measured	& Calculated	& Measured	& Calculated	\\ \hline  
$\rm ^4I_{15/2}$ 1	& 0			& 0		    	& 0			& 0	      \\                     
2				& 25.2 (26.2)	& 25.25	    	& 41.1 (41.1)	& 42.6    \\
3				& 75.1		& 75.3	    	& 74.6 (74.6)	& 68.3    \\           
4				& 98.8 (97.3)	& 97.3	    	& 92.4 (93.4)	& 94.2    \\
5				& 111.2 (111.7)	& 109.2	    	& 162.2 (162.2)	& 165.4  \\ 
6				& 463.5		& 457	    	& 431.8		& 427.6  \\                     
7				& 491.6 (491.6)	& 486.1	    	& 453.6 (453.6)	& 457.3  \\ 
8				& 506		& 497.4	    	& 533.6		& 528.4  \\ \hline                     				                                         
$\rm ^4I_{13/2}$ 1	& 6503.8	    	& 6504.3		& 6517.8	    	& 6517.7\\
2				& 6540.5	    	& 6540.2		& 6568.1	    	& 6560.6\\        
3				& 6556		& 6556		& 6579.3	    	& 6575.8\\            
4				& 6580.5	    	& 6576.3		& 6635.1	    	& 6632.3\\        
5				& 6826.3	    	& 6829.4		& 6823.9	    	& 6815.9\\        
6				& 6842.1	    	& 6838.4		& 6831.6	    	& 6822.9\\        
7				& 6850.8	    	& 6844		& 6917.9	    	& 6901.1\\ \hline                                      
$\rm ^4I_{11/2}$ 1	& 10196		& 10197		& 10192		& 10193 \\ 
2				& 10219		& 10219		& 10231		& 10230 \\           
3				& 10232		& 10231		& 10258		& 10259 \\           
4				& 10342		& 10349		& 10336		& 10343 \\           
5				& 10348		& 10354		& 10340		& 10346 \\           
6				& 10355		& 10359		& 10394		& 10405 \\ \hline           
$\rm ^4I_{9/2}$  1	& 12331.7	    	& 12332		& 12343		& 12345 \\
2				& 12404.9	    	& 12400		& 12430		& 12425 \\         
3				& 12524.6	    	& 12531		& 12533		& 12539 \\         
4				& 12581.4	    	& 12584		& 12584		& 12599 \\         
5				& 12621		& 12621		& 12622		& 12630 \\ \hline            
$\rm ^4F_{9/2}$  1	& 15129.1	    	& 15132		& 15138.6	    	& 15142 \\
2				& 15156.1	    	& 15156		& 15176.6	    	& 15186 \\       
3				& 15301.4	    	& 15308		& 15276.7	    	& 15284 \\       
4				& 15333.8	    	& 15340		& 15377.9	    	& 15386 \\       
5				& 15446.8	    	& 15440		& 15457.4	    	& 15457 \\ \hline       
$\rm ^4S_{3/2}$  1	& 18184.8	    	& 18190		& 18201		& 18205 \\
2				& 18234.8	    	& 18235		& 18295		& 18295 \\ \hline                
$\rm ^2H_{11/2}$ 1	& 19048.1	    	& 19052		& 19054.1	    	& 19057 \\
2				& 19066.3	    	& 19063		& 19072.1	    	& 19079 \\       
3				& 19074.5	    	& 19076		& 19088.5	    	& 19104 \\       
4				& 19193		& 19207		& 19205		& 19213 \\           
5				& 19209.5	    	& 19228		& 19229		& 19242 \\         
6				& 19265.2	    	& 19271		& 19273.7	    	& 19274 \\ \hline          
$\rm ^4F_{7/2}$  1	& 20379		& 20388		& 20358.8	    	& 20351 \\
2				& 20437.5	    	& 20438		& 20462.2	    	& 20452 \\       
3				& 20478.8	    	& 20484		& 20493.8	    	& 20493 \\       
4				& 20543.9	    	& 20535		& 20585.8	    	& 20575 \\ \hline           
$\rm ^4F_{5/2}$  1	& 22047.2	    	& 22063		& 22053.5	    	& 22061 \\
2				& 22070		& 22078		& 22087.5	    	& 22100 \\         
3				& 22194.7	    	& 22179		& 22193.8	    	& 22182 \\ \hline            
$\rm ^4F_{3/2}$  1	& 22451		& 22453		& 22428.8	    	& 22428 \\
2				& 22514.1	    	& 22514		& 22570.5	    	& 22566 \\ \hline       
\end{tabular}
\end{center}
\label{TableII}
\end{table}

Crystal field energies for both the R1 and R2 centers obtained from the analysis of the experimental data are presented in Table~\ref{TableII}.
The obtained energy level patterns, and the total splittings of the ground and excited multiplets are similar to those of the impurity Er$^{3+}$ions at the Y$^{3+}$ sites with C$_2$ symmetry in $\rm Y_2O_3$~\cite{Chang_1982}.
The energy level pattern of the Er$^{3+}$ground multiplet $\rm ^4I_{15/2}$ in the octahedral CF contains two groups of sublevels, separated by the energy gap of about 300~cm$^{-1}$ in the case of oxygen ligands, the lower one involves a quadruplet $\Gamma_8$, a doublet $\Gamma _6$ and a second quadruplet $\Gamma_8$, and the upper one involves a doublet $\Gamma_7$ and a quadruplet $\Gamma_8$, the total spitting being close to 500~cm$^{-1}$.
Due to a splitting of the quadruplets by the low-symmetry component of the crystal field, one may expect to observe a pattern formed by the lower five doublets well separated from the upper three doublets.
The measured CF energies for both the Er1 and Er2 centers agree well with these simple arguments.

The two optical centers formed by the Er$^{3+}$ions in \syo\ differ markedly in the splittings of the $\rm ^4S_{3/2}$ and $\rm ^4F_{3/2}$ multiplets (see Table~\ref{TableII}).
Large splittings of the multiplets with the total angular moment $J=3/2$ are likely be related to the more pronounced distortion of the first coordination shell (oxygen octahedron) for the Er$^{3+}$ions at the R2 sites and, consequently, stronger quadrupole components of the crystal field.
This conjecture is confirmed below by the CF calculations (see Sec.~\ref{subsec_CF}).

\subsection{Inelastic neutron scattering}

\begin{figure}[tb] 
\centering
\includegraphics[width=0.88\columnwidth]{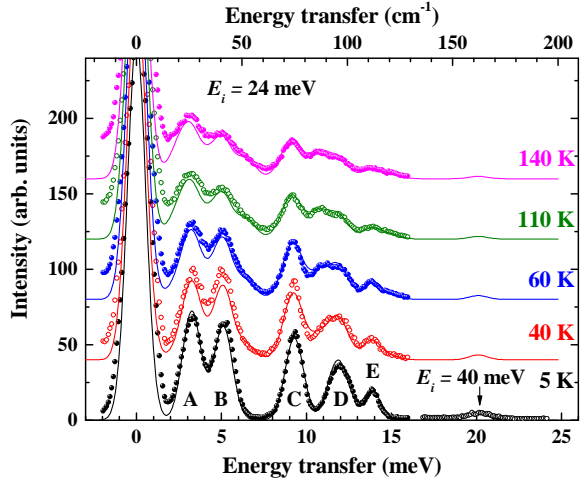}
\caption{(Color online) Neutron scattering spectra of a polycrystalline \seo\ sample for the incident neutron energy $E_i=24$~meV measured at different temperatures.
The arrow points to the additional peak in the scattering intensity observed with the incident energy of 40~meV.
The origin of the peaks labeled from A to E is discussed in the text.
The results of simulations are represented by solid lines. 
The curves at different temperatures are offset for clarity.}
\label{Fig6_INS}
\end{figure}

The results of the inelastic neutron scattering measurements on a powder sample of \seo\ are summarised in Fig.~\ref{Fig6_INS}.
At a base temperature of 5~K, several peaks are clearly visible in the energy transfer range of up to 15~meV, they are labeled as A, B, C, D, and E.
The widths of the peaks are resolution limited, apart from peak D, which consists of two excitation levels (see Sec.~\ref{subsec_INS}).
The peaks show very little dependence (a slight decrease) of intensity with the scattering vector $Q$, consistent with the $Q$-dependence of the magnetic form factor for the Er$^{3+}$ions.
There were no detectable phonon signals observed at any $Q$, therefore the intensity of the peaks is presumed to be fully magnetic in nature.

With increasing temperature, the main peaks start to look broader and several weaker peaks become visible at different energy transfers.
This behavior is typically associated with the increase of the fractional population of the higher-energy levels at higher temperatures, which allows excited state transitions to become visible.

Additional measurements with a higher incident neutron energy revealed the presence of a much weaker peak at about 20~meV (see Fig.~\ref{Fig6_INS}) as well as a weak double peak at about 60~meV (not shown).
  
\subsection{Magnetometry studies}
\begin{figure}[b] 
\centering
\includegraphics[width=0.9\columnwidth]{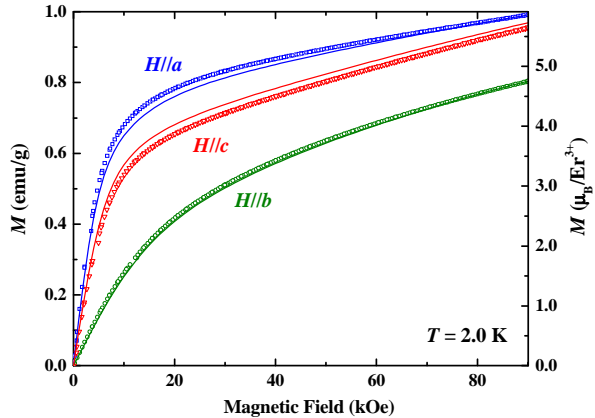}
\caption{(Color online) Measured (symbols) and calculated (lines) magnetic field dependencies of the magnetization along the $a$, $b$ and $c$ crystallographic directions in the \syoEhalf\ sample at $T=2.0$~K.}
\label{Fig7_M_Y}
\end{figure}

\begin{figure}[tb] 
\centering
\includegraphics[width=0.9\columnwidth]{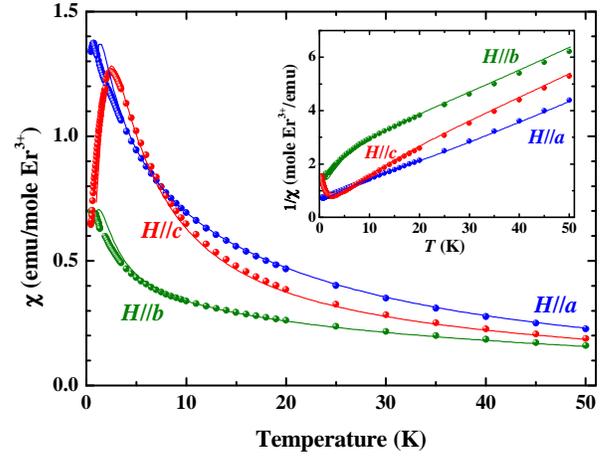}
\caption{(Color online) Temperature dependence of dc magnetic susceptibility of \seo\ crystal measured along the three crystallographic axes (symbols) in a field of $H=1$~kOe in the temperature range of 0.5 -- 50~K~\cite{Hayes_2012,Balakrishnan_2009}.
Solid lines represent the calculated susceptibility for temperatures from 1 to 50~K in the framework of the four-particle self-consistent model (see Sec.~\ref{sec_4p_model} for further details).
Inset: the temperature dependence of the inverse susceptibility.}
\label{Fig8_chi}
\end{figure}

\begin{figure}[tb] 
\centering
\includegraphics[width=0.9\columnwidth]{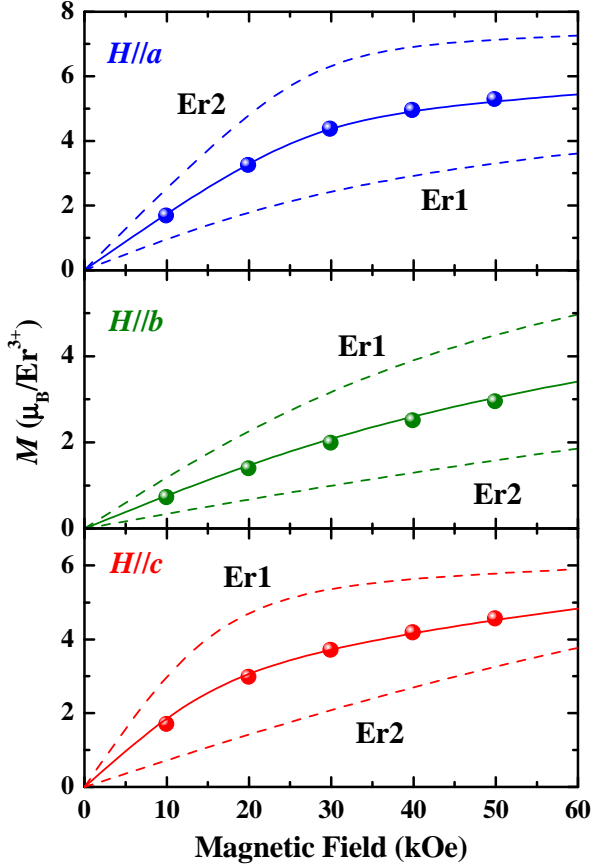}
\caption{(Color online) Measured (symbols) and calculated (solid lines) field dependence of the magnetization of \seo\ single-crystals along the three crystallographic axes at $T=5$~K.
The calculated magnetic moments at sites R1 and R2 are shown by dashed lines marked as Er1 and Er2, respectively.}
\label{Fig9_M_Er}
\end{figure}

Magnetization results at $T=2$~K for a diluted \syoEhalf\ sample are shown in Fig.~\ref{Fig7_M_Y}, while Figures~\ref{Fig8_chi} and \ref{Fig9_M_Er} summarize the results of the susceptibility and magnetization measurements in the \seo\ single crystals.

For the diluted sample, the magnetization increases in an applied field in a way which preserves the order of the absolute values of magnetization observed in a concentrated sample at higher temperatures, $M_{H \parallel a} > M_{H \parallel c} > M_{H \parallel b}$.
The gradients of all three curves gradually decrease with an applied field, as would be expected for a system of noninteracting magnetic moments.
The details of the magnetization calculations for a diluted sample are given in Sec.~\ref{subsec_CF}.

In Fig.~\ref{Fig8_chi} we combine the previously reported results for the temperature dependence of the susceptibility of \seo~\cite{Hayes_2012,Balakrishnan_2009} with the results of the calculations.
For all three directions of an applied magnetic field, the four-particle model described in detail in Sec.~\ref{sec_4p_model} returned a satisfactory agreement with the experimental data, particularly at higher temperatures.
A linear temperature dependence of the inverse susceptibilities along the three crystallographic axes is observed at $T>15$~K (see inset in Fig.~\ref{Fig8_chi}).
The negative values of the corresponding Curie-Weiss temperatures~\cite{Hayes_2012} indicate \afm\ interactions between the Er$^{3+}$ ions.

Fig.~\ref{Fig9_M_Er} shows a comparison between the experimental and the theoretical results for the magnetization of \seo\ at $T=5$~K.
Although the $M(H)$ curves have been reported previously, in particular in high fields~\cite{Petrenko_2008} and at very low temperatures~\cite{Hayes_2012}, comparison with the more accurate experimental data helps to justify the validity of the  theoretical approach used (see Sec.~\ref{sec_4p_model} for further details).
One important observation to make here is that the magnetization for $\bm{H} \parallel a$ is largely due to the Er$^{3+}$ ions at the R2 sites, while for $\bm{H} \parallel b$ and $\bm{H} \parallel c$, the dominant contribution is from the Er$^{3+}$ ions at the R1 sites. 
Combined with the results of polarized neutron scattering~\cite{Hayes_2011}, which suggested that at low temperatures the magnetic moments in the long-range and short-range structures of \seo\ are predominantly pointing along the $c$ and $a$ axes, respectively, this observation uniquely identifies the R1 sites with the commensurate long-range order and the R2 sites with the incommensurate short-range order.
 
\section{Modeling and Discussion}
\label{sec_modeling}
\subsection{Crystal fields affecting the Er$^{3+}$ ions at the R1 and R2 sites}
\label{subsec_CF}

The analysis of the experimental data presented in Sec.~\ref{sec_results} was carried out starting from the Hamiltonian, ${\cal H}$, of a single Er$^{3+}$ ion operating in the total space of 364 states of the ground electronic $4f^{11}$ configuration:
\begin{equation}
{\cal H} = {\cal H}_{\rm FI} + {\cal H}_{\rm CF}.
\label{eq7_ham}
\end{equation}
Here
\begin{eqnarray}
\nonumber
{\cal H}_{\rm FI} & = & \zeta \sum_j \hat{\bm{l}}_j \hat{\bm{s}}_j + \alpha \hat{\bm{L}}^2 + \beta \hat{G}(G_2) + \gamma \hat{G}(R_7) \\* 
                          & + & \sum_q (F^q \hat{f}_q + P^q \hat{p}_q + T^q \hat{t}_q + M^q \hat{m}_q)
\label{eq8_ham-fi}
\end{eqnarray}
is the free-ion Hamiltonian written in a standard form that includes the energies of the spin-orbit interaction, electrostatic interactions between the $4f$ electrons (labeled by the index $j$), and additional terms due to interconfiguration and relativistic interactions, $\hat{\bm{l}} _j$ and $\hat{\bm{s}} _j$ are the orbital and spin moments of the electrons, $\bm{L}$ is the total orbital moment, explicit expressions for the operators $\hat{G}, \hat{f}, \hat{p}, \hat{t}, \hat{m}$ are given in literature (see Ref.~\cite{Carnall_1989}). 
In the present work, we use parameters of the Hamiltonian~(\ref{eq8_ham-fi}) from Ref.~\cite{Magnani_2002} for the Er$^{3+}$ impurity ions in $\rm BaY_2F_8$ modified slightly to fit the gaps between the multiplet centers of gravity.
Namely, the same parameters $F^2=96320$, $F^4=67854$, $F^6=54260$, $\alpha=18.2$, $\beta=-600$, $\gamma=1760$, $P^2=594$, $P^4=445$, $P^6=119$, $M^0=3.7$, $M^2=2.07$, $M^4=1.15$, $T^2=390$, $T^3=50$, $T^4=83$, $T^6=-271$, $T^7=298$, $T^8=280$~(cm$^{-1}$) for both the R1 and R2 sites, but different spin-orbit coupling constants $\zeta=2361$ and 2366~cm$^{-1}$ are used for the R1 and R2 sites, respectively.

The Hamiltonian ${\cal H}_{\rm CF}$ gives the energy of the localized $4f$ electrons in a static crystal field.
In the crystallographic coordinate system with the $z$~axis along the $c$~axis, this Hamiltonian ${\cal H}_{\rm CF}$ for a RE ion in the crystal field with the $C_s$ symmetry in \sRo, is determined by 15~CF parameters $B^q_p$:
\begin{eqnarray}
\label{eq9_ham-cf}
{\cal H}_{\rm CF}	& =	& B^0_2 O^0_2 + B^2_2 O^2_2 + B^{-2}_2 O^{-2}_2 + B^0_4 O^0_4 + B^2_4 O^2_4 \\* \nonumber 
				& +	& B^{-2}_4 O^{-2}_4 + B^4_4 O^4_4 + B^{-4}_4 O^{-4}_4 + B^0_6 O^0_6 + B^2_6 O^2_6 \\* \nonumber 
				& +	& B^{-2}_6 O^{-2}_6 + B^4_6 O^4_6 + B^{-4}_6 O^{-4}_6 + B^6_6 O^6_6 + B^{-6}_6 O^{-6}_6.
\end{eqnarray}
Here $O^q_p$ are linear combinations of the spherical tensors $C^{(p)}_q$~\cite{Klekovkina_2011,note_eq8_9}. 
In particular, $O_p^0 = a_{p0} C_0^{(p)}$, $O_p^q = a_{pq} (C_q^{(p)} + C_{-q}^{(p)})$, $O_p^{-q} = -ia_{pq}(C_q^{(p)}-C_{-q}^{(p)})$ for $q=2,4,6$; the numerical factors are $a_{20}=2$, $a_{22}=2/\sqrt{6}$, $a_{40}=8$, $a_{42}=4/\sqrt{10}$, $a_{44}=8/\sqrt{70}$, $a_{60}=16$, $a_{62}=16/\sqrt{105}$, $a_{64}=16/\sqrt{126}$, $a_{66}=16/\sqrt{231}$. 
There are two sets of independent CF parameters for the RE ions at the nonequivalent R1 and R2 positions, parameters $B_p^{-q} (q > 0)$ for the ions in the sublattices 1, 3 and 2, 4 (5, 7 and 6, 8) have the same absolute values but differ in sign.

The initial values of the CF parameters were calculated in the framework of the semi-phenomenological Exchange Charge Model (ECM)~\cite{Malkin_1987} using the expressions
\begin{eqnarray}
\nonumber
B_p^q = K_p^q \sum_\nu \frac{e^2}{R_\nu} \bigg[-Q_\nu (1-\sigma_p) \frac{\langle r^p \rangle}{{R_\nu^p}} \hspace{15mm} \\
+ \frac{2(2p+1)}{7} S_p(R_\nu)\bigg] O_p^q (\Theta_\nu, \Phi_\nu),
\label{eq10_Bqp}
\end{eqnarray}
where $K_p^0 = a_{p0}^{-2}$, $K_p^q = a_{pq}^{-2}/2$ for $q\neq 0$, $e$ is the elementary charge, $R_\nu$, $\Theta _\nu$, $\Phi _\nu$ are spherical coordinates of the ion located at the lattice site $\nu$, $eQ_\nu$ are ion charges, $\sigma_p$ are the shielding factors, and $\langle r^p \rangle$ are the moments of the $4f$ electron density.
Linear combinations $S_p (R_\nu)$ of the squared overlap integrals between the wave functions of the $4f$ electrons localized at the RE ion and the electronic wave functions of the lattice ions determine the ``exchange" charges. 
Because the overlap integrals decrease exponentially with increasing distance between the ions, we take into account only the overlapping between the Er$^{3+}$ $\vert 4fl_z \rangle$ wave functions and the $\vert 2s \rangle$, $ \vert 2pl_z \rangle$ wave functions corresponding to the outer electronic shells of the six nearest neighbor oxygen ions (here $l_z$ is the projection of the electronic orbital moment on the bond direction). 
In this case, introducing the dimensionless model parameters $G_s$, $G_\sigma$ and $G_\pi$, we obtain~\cite{Malkin_1987}
\begin{eqnarray}
\nonumber
S_p (R_\nu)	& =	& G_s S_s (R_\nu)^2 + G_\sigma S_\sigma (R_\nu)^2 \\*
			& +	& [2-\frac{p(p+1)}{12}] \: G_\pi S_\pi (R_\nu)^2.
\label{eq11_SpRv}
\end{eqnarray}
The $R$ dependencies of the overlap integrals $S_s (R) = \langle 4f0\vert 2s \rangle$, $S_\sigma (R) = \langle 4f0 \vert 2p0 \rangle$ and $S_\pi (R) = \langle 4f1 \vert 2p1\rangle$, computed with the available radial wave functions of the Er$^{3+}$~\cite{Freeman_1962} and O$^{2-}$~\cite{Clementi_1964} ions, can be approximated by the functions $S_u = a_u \exp (-b_u R^{c_u})$ ($u=s, \sigma$ and $\pi$), where $a_s=0.333$, $b_s=1.0776$, $c_s=1.4269$; $a_\sigma = 0.08648$, $b_\sigma =0.3845$, $c_\sigma =1.9254$; $a_\pi = 3.1067$, $b_\pi = 3.0381$, $c_\pi = 0.8048$ ($R$ in angstroms). 
The results of the calculations of the CF parameters~(\ref{eq10_Bqp}) performed in the framework of the simplest single-parameter version of ECM ($G_s = G_\sigma = G_\pi = 8$) are presented in Table~\ref{TableIII}. 
The values of $\langle r^2 \rangle = 0.666$, $\langle r^4 \rangle =1.126$ and $\langle r^6 \rangle =3.978$ (in atomic units)~\cite{Freeman_1962}, $\sigma _2 = 0.617$~\cite{Gupta_1973}, $\sigma_4=\sigma_6=0$, the nominal values of the ion charges $Q$(Sr)=2, $Q$(Y)=3, $Q$(O)=$-2$ and the structure parameters of \seo~\cite{Li_2014_SEO} were used in the calculations. 
It should be noted that the oxygen coordinates in the unit cell of \syo\ determined in Refs.~\cite{Muller-Buschbaum_1968,Fu_2005} should be revised because they lead to significantly overestimated distances between the Y$^{3+}$ and the nearest neighbor O$^{2-}$ ions as compared to the corresponding distances in \seo.
In particular, according to Ref.~\cite{Fu_2005}, the average Y$^{3+}$--O$^{2-}$ distances in Y1O$_6$ and Y2O$_6$ octahedra exceed the corresponding average Er$^{3+}$--O$^{2-}$ distances in \seo\ by 0.08~\AA.

\begin{table}[tb] 
\caption{Crystal field parameters $B^q_p$ (cm$^{-1}$) for the Er$^{3+}$ ions at $r_{1,{\rm R}1}$ and $r_{1,{\rm R}2}$ sites ($Calc.$ -- the results of calculations in the framework of the exchange charge model, $Fin.$ -- the final values from the fitting procedure).}
\begin{center}
\begin{tabular}{|l|r|c|c|c|c|}
\hline
\hline
\multicolumn{2}{|c|}{}		&  \multicolumn{2}{|c|}{R1}& \multicolumn{2}{|c|}{R2} \\ \hline	                                               
$p$ & $q$				& $Calc.$	& $Fin.$		& $Calc.$		& $Fin.$     \\ \hline                                                                                   
2	&	0	& 118.7	& 188		& -33.1		& 17        \\                                                     
2	&	2	& 118.6	& 137.5		& -546.9		& -744     \\                                                   
2  	&	-2	& -104.2	& -171.2		& -110.8		& -125     \\                                                 
4	&	0	& -59.8	& -57.3		& -55.9		& -60.2    \\                                                   
4	&	2	& -1012	& -1066.2		& 1064		& 1033.2  \\       
4  	&	-2	& 1140	& 1165.2		& -797.8		& -977.8   \\      
4	&	4	& -73.3	& -86.9		& 379		& 430.2    \\         
4   	&	-4	& -1015	& -972.3		& -638.3		& -685.6   \\     
6	&	0	& -39.5	& -38		& -33.0		& -35.2     \\           
6	&	2	& -8.2	& -22.3		& -51.6		& -68.4     \\        
6   	&	-2	& 23.0	& 22.8		& -13.6		& -42.8     \\         
6	&	4	& -8.6	& 30.1		& -63.4		& -80.2     \\         
6   	&	-4	& -145.6	& -115.2		& -186.8		& -191.4   \\     
6	&	6	& -208	& -162.2		& -66.7		& -119.6   \\        
6	& 	-6	& -147.6	& -84		& 56.3		& 80.5      \\ \hline
\end{tabular}
\end{center}
\label{TableIII}
\end{table}    

The energy level schemes and the ratios between the $g$-tensor components of the ground doublet for the Er$^{3+}$ ions at the R1 and R2 sites obtained from the numerical diagonalization of the Hamiltonian~(\ref{eq7_ham}) which make use of the calculated CF parameters agree only qualitatively with the experimental data.
In particular, the splittings of the $\rm ^4S_{3/2}$ and $\rm ^4F_{3/2}$ multiplets are underestimated by about 30\%.

The next step in the simulations involves variations of the initial values of the CF parameters. 
To conserve the physical meaning of the CF parameters during the fitting procedure, namely, their definition within the crystallographic coordinate system, the CF parameters were varied to match simultaneously not only the measured energies of the CF levels (see Table~\ref{TableII}) but the principal values and orientations of the principal axes of the $g$-tensor of the ground doublet of the Er$^{3+}$ ion (see Table~\ref{TableI}) as well.

When a strongly diluted paramagnetic crystal \syoE\ is placed in an external magnetic field $\bm{H}$, the Hamiltonian~(\ref{eq7_ham}) of an Er$^{3+}$ ion contains an additional term, the Zeeman energy, ${\cal H}_Z = - \bm{\hat{m} H}$. 
Here $\bm{\hat m} = -\mu_{\mathrm B} \sum_j (k\bm{\hat l}_j + 2 \bm{\hat s}_j)$ is the magnetic moment operator, the sum is taken over eleven $4f$ electrons, and $k$ is the orbital reduction factor that, for simplicity, is approximated by a scalar. 
The projection of the Zeeman energy on the 2-dimensional space of wave functions ($\sigma =+$ and $-$) of a Kramers doublet $\Gamma$ can be written as ${\cal H}_S = \mu_{\mathrm B}\sum\limits_{\alpha \beta } H_\alpha g_{\alpha \beta} \hat{S}_\beta$,
where the effective spin $S=1/2$, and components of the $g$-tensor are determined by the corresponding matrix elements of the magnetic moment:
\begin{eqnarray}
\nonumber
g_{\alpha x} & = & 2\operatorname{Re}\langle \Gamma, + \vert \hat{m}_\alpha \vert \Gamma, -\rangle, \\* \nonumber
g_{\alpha y} & = &-2\operatorname{Im}\langle\Gamma, + \vert \hat{m}_\alpha \vert \Gamma, -\rangle, \\*
g_{\alpha z} & = & 2\langle\Gamma, + \vert \hat{m}_\alpha \vert \Gamma, + \rangle.
\label{eq12_gxyz}
\end{eqnarray}
Square roots of eigenvalues of the matrix $(g^2)_{\alpha\beta}=\sum\limits_\gamma g_{\alpha \gamma } g_{\beta \gamma}$ are the principal values of the $g$-tensor, and eigenvectors of this matrix are the directional cosines of the corresponding principal axes. 
In the present work we considered the orbital reduction factor as an additional fitting parameter with the condition that it may only deviate from unity by a small amount. 
The fitting procedure involved numerical diagonalization of the Hamiltonian~(\ref{eq7_ham}) for each fixed set of the CF parameters and subsequent computations of the $g$-tensor components~(\ref{eq12_gxyz}) of the ground doublet;
the main attention was paid to the correct description of the CF energies of the ground multiplet sublevels and principal values of the $g$-tensor. 
It is well established that the conventional one-electron CF parameterization underestimates the splitting of the $\rm ^2H_{11/2}$ multiplet of Er$^{3+}$~\cite{Jayasankar_1989,Moune_1991,Gruber_1993,Tanner_2002}. 
Following Refs.~\cite{Moune_1991} and~\cite{Tanner_2002}, to fit the measured splittings, we introduced additional empirical parameters, the multiplying factors of 1.37 and 1.32 for the Er$^{3+}$ ions at the R1 and R2 sites, respectively, in the reduced matrix elements of the fourth-rank spherical tensors in the space of states belonging to this multiplet. 
The final values of the CF parameters are presented in Table~\ref{TableIII}, and the calculated $g$-factors and the CF energies are compared with the experimental data in Tables~\ref{TableI} and~\ref{TableII}, respectively. 
Using the obtained sets of the CF parameters, we are also able to satisfactorily reproduce the results of the magnetization measurements. 
From a comparison of the calculated and measured field dependencies of the magnetization along the crystallographic axes at a temperature of 2~K for the \syoEhalf\ sample assuming different populations $c_\kappa$ of the Er$^{3+}$ ions at the R1 and R2 sites (see Fig.~\ref{Fig7_M_Y}), we obtained physically meaningful values of $c_{\rm R1}=57$\% and $c_{\rm R2}=43$\%. 
This result proves the preferential occupation of the R1 sites by the impurity Er$^{3+}$ ions in the strongly diluted paramagnet \syoE.

\subsection{Inelastic neutron scattering}
\label{subsec_INS}
Intensity of neutron scattering from powder \seo\ is considered here as a sum of two terms (see Eq.~\ref{eq13_iet} below) corresponding to elastic and inelastic scattering. 
Neglecting magnetic interactions between the paramagnetic ions, \textit{i.e.}, supposing that the Er$^{3+}$ ions contribute independently to the scattering processes, we can write the scattering intensity versus energy transferred from the incident neutrons to the magnetic subsystem, averaged over directions of the scattering vector, as follows~\cite{Trammell_1953}
\begin{eqnarray}
\label{eq13_iet}
I(\Delta E,T) & = & K_{f}D(0,\Delta E) \\*  \nonumber
		   & +  & \sum \limits_{\kappa \Gamma \Gamma ^\prime} p_{\kappa ,\Gamma }(T)D(E_{\kappa \Gamma^\prime} - E_{\kappa \Gamma},\Delta E) w_{\kappa}(\Gamma,\Gamma^\prime),
\end{eqnarray}
where
\begin{equation}
D(x,\Delta E) = (2\pi \delta ^2)^{-1/2}\exp [-(x-\Delta E)^2/2\delta^2] 
\label{eq14_dxe}
\end{equation}
is the setup form-function that is approximated by a Gaussian, $p_{\kappa,\Gamma }(T)$ are populations of energy levels $E_{\kappa \Gamma}$ with wave functions $\vert \kappa \Gamma,\sigma \rangle$ of Er$^{3+}$ ions at $\kappa$-sites, $\kappa = $Er1, Er2;
\begin{equation}
w_{\kappa }(\Gamma ,\Gamma^\prime) = \sum \limits_{\sigma\sigma^\prime,~\alpha =x,y,z}\vert\langle\kappa \Gamma^\prime, \sigma^\prime | \hat{m}_{\alpha} |\kappa \Gamma ,\sigma \rangle\vert^2
\label{eq15_dipoles}
\end{equation}
are relative probabilities of magnetic dipole transitions between energy levels of a RE ion, and $K_f$ is a fitting parameter. 
Expression~(\ref{eq15_dipoles}) is a reasonable approximation for the scattering intensity assuming equal probabilities for the directions of the crystallographic axes in a powder sample.
The spectral envelopes~(\ref{eq13_iet}) were calculated by making use of the transition energies given in Table~\ref{TableII}, the transition probabilities computed with the wave functions obtained from the numerical diagonalization of Hamiltonian~(\ref{eq7_ham}) for impurity Er$^{3+}$ ions in \syo, and a $\delta$ value of 4.4~cm$^{-1}$.
The corresponding HWHM of $\sqrt{2 \ln{2}} \: \delta = 0.65$~meV of the spectral lines is chosen to coincide with the average instrumental resolution for the neutrons with an incident energy of 24~meV.
The results are compared with the experimental data in Fig.~\ref{Fig6_INS}.

At low temperatures, in particular, at 5~K, the spectrum is formed by transitions of the Er$^{3+}$ ions from the ground doublet to the excited crystal field sublevels of the $\rm ^4I_{15/2}$ multiplet. 
According to the calculations, the peaks A and B (see Fig.~\ref{Fig6_INS}) correspond to the transitions with integrated intensities (in units of the intensity of the transition A at the R1 sites) of 1.0 and 0.86 to the first excited doublets with energies 26.2 and 41.1~cm$^{-1}$ at the R1 and R2 sites, respectively. 
Both the peaks C and D correspond to the unresolved closely spaced transitions to the third (with energies 75.1 and 74.6~cm$^{-1}$) and fourth (with energies 97.3 and 93.4~cm$^{-1}$) sublevels of the Er$^{3+}$ ions at the R1 and R2 sites, with the total intensities of $0.45+0.30=0.75$ and $0.35+0.26=0.61$, respectively.
The peak E with an intensity of 0.264 corresponds to transitions from the ground state to the fifth sublevel of the $\rm ^4I_{15/2}$ multiplet with an energy of 111.7~cm$^{-1}$ at the R1 sites. 
Note that the calculated intensities of transitions from the ground state to the upper three sublevels of the $\rm ^4I_{15/2}$ multiplet at both the R1 and R2 sites are about two orders of magnitude weaker than the intensities of transitions within the lower groups of five sublevels.
The temperature evolution of the spectrum envelope is caused by the redistribution of the sublevel populations, a number of additional transitions appear which lead to the smoothing of the scattering spectra at higher temperatures.

\subsection{Four-particle self-consistent model: dc magnetic susceptibility and the magnetization}
\label{sec_4p_model}
The above discussions prove that the Er$^{3+}$ ions at the R1 and R2 sites have rather different magnetic anisotropy in the ground state, the easy axis of the magnetization is parallel to the chain direction (along the $c$~axis) for Er1, and is perpendicular to the chain direction for Er2. 
At temperatures of about 1~K the dipole-dipole interactions tend to induce a long-range ferromagnetic order along the chains at the Er1 sites or an \afm\ order of the N\'{e}el type at the Er2 sites because of small distances between the nearest neighbor ions in the chains and large values of the $g$-factors $g_3$ (Er1) and $g_1$ (Er2).
In particular, when considering only the dipolar interactions, simulations of the dc susceptibility along the $c$~axis in the framework of the conventional single-site mean-field approximation bring about a divergence at the Curie-Weiss temperature of 2.1~K. 
Powder neutron diffraction and single-crystal heat capacity measurements have demonstrated long-range magnetic ordering in \seo\ below $T_{\rm N}=0.75$~K, but with a rather specific magnetic structure: the magnetic moments at the Er1 sites point along the $c$~axis and form ferromagnetic chains along this direction, with neighboring chains coupled antiferromagnetically~\cite{Petrenko_2008}. 
Note that the measured value of the magnetic moment at the Er1 sites of 4.5$\mu_{\mathrm B}$ at $T=0.55$~K is in line with the corresponding $g$-factor $g_3=9.93$ which is determined in the present work. 
The measured magnetic moments of the Er$^{3+}$ ions at the R2 sites below $T_{\rm N}$ have much reduced values, and it has been shown that only short-range magnetic correlations emerge within the R2 chains coexisting with the long-range order in the R1 chains~\cite{Petrenko_2008,Hayes_2011}.

In the following we focus our attention on the magnetic properties of \seo\ in the paramagnetic phase.
Each Er$^{3+}$ ion (at both the R1 and R2 sites) has eight neighboring Er$^{3+}$ ions at distances less than 0.4~nm. 
For example, an ion in the sublattice number 1 (at the R1 site) has the two nearest neighbor ions belonging to the same sublattice (to the left and right within the chain, see Fig.~\ref{Fig1_structure}), a pair of the next nearest neighbor R1 ions belonging to the magnetically equivalent sublattice number 3 (on the second leg of the ladder), and two pairs of more distant ions at the R2 sites belonging to the sublattices with numbers 6 and 7. 
There are good reasons to believe that the neighboring ions are connected by strong \afm\ exchange interactions, such as the observed \afm\ ordering of the R1 chains, and the measured temperature dependence of the susceptibility (see Fig.~\ref{Fig8_chi}), which demonstrates the suppressed responses of the Er$^{3+}$ ions in external magnetic fields at low temperatures.

To elucidate the temperature dependence of the dc susceptibility (in particular, the broad peak observed in the low-temperature susceptibility along the $c$~axis), and the field dependence of the isothermal magnetization, we employ the well known Bethe-Peierls approximation and introduce the four-particle self-consistent model. 
In the framework of this model we treat the dipole-dipole interactions exactly by computing the lattice sums with the Ewald method and consider the exchange coupling constants as fitting parameters.

Let us consider four neighboring Er$^{3+}$ ions in a zig-zag chain labeled sequentially by indices I, II, III and IV (as shown in Fig.~\ref{Fig1_structure}). 
The effective Hamiltonian of this cluster has the following form:
\begin{equation}
{\cal H}_C = {\cal H}_{C0} - (\bm{\hat m}_{\rm I}+\bm{\hat m}_{\rm IV}) \bm{H}_{loc 1} - (\bm{\hat m}_{\rm II} + \bm{\hat m}_{\rm III}) \bm{H}_{loc 2} ,
\label{eq16_ham-c}
\end{equation}
where $\bm{H}_{loc 1}$ and $\bm{H}_{loc 2}$ are local magnetic fields affecting the external and internal pairs of the ions (these two pairs of the ions -- the first and the fourth, the second and the third -- are not equivalent in the construction). 
The Hamiltonian $\mathcal{H}_{C0}$ involves the single-ion energies $\mathcal{H}_j$ and the energies of interactions between the ions in the cluster:
\begin{eqnarray}
\nonumber
{\cal H}_{C0} 	& = & \sum\limits_{j={\rm I, IV}} {\cal H}_j + \bm{\hat m}_{\rm I} \bm{J}_r \bm{\hat m}_{\rm II} + \bm{\hat m}_{\rm II} \bm{J}_r^\prime \bm{\hat m}_{\rm III} \\* 
			& + & \bm{\hat m}_{\rm III} \bm{J}_r \bm{\hat m}_{\rm IV} + \bm{\hat m}_{\rm I} \bm{J}_c \bm{\hat m}_{\rm III} + \bm{\hat m}_{\rm II} \bm{J}_c \bm{\hat m}_{\rm IV} .
\label{eq17_ham-c0}
\end{eqnarray}
The matrices $\bm{J}_r = \bm{J}_r^{(dd)} + \bm{J}_r^{(ex)}$, $\bm{J}_r^\prime = \bm{J}_r^{\prime (dd)}+\bm{J}_r^{\prime (ex)}$ and $\bm{J}_c = \bm{J}_c^{(dd)} + \bm{J}_c^{(ex)}$ determine the dipole-dipole ($dd$) and the exchange ($ex$) interactions between the Er$^{3+}$ ions along the rungs, $r$, and legs, $c$, of the ladder under consideration.
The operators (\ref{eq16_ham-c}) and (\ref{eq17_ham-c0}) work in the Kronecker product of the four single-ion Hilbert spaces of states. 
For magnetic fields of less than 60~kOe and temperatures below 60~K, computations of the magnetization and susceptibility of the single Er$^{3+}$ ion at the R1 (or R2) site have shown that results obtained with the Hamiltonian determined in the total space of 364 states of the electronic $4f^{11}$ configuration, and those with the single-ion Hamiltonian determined in the truncated Hilbert space spanned by the wave functions of the three lowest CF doublets, do not differ by more than a few percent.
Based on these results, we constructed the matrix representation of the Hamiltonian (\ref{eq16_ham-c}) in the basis of $6^4=1296$ states corresponding to the three lower sublevels of the ground multiplet of each of the four Er$^{3+}$ ions in the cluster. 
The contributions to the magnetic moments due to other excited single-ion states were taken into account in the last step of the simulations by adding the corresponding differences estimated from the exact single-ion calculations.

To model the temperature dependence of the dc susceptibility, first we compute the responses of the Er$^{3+}$ ions in the cluster to weak local magnetic fields. 
For a given set of the exchange coupling constants in the Hamiltonian (\ref{eq17_ham-c0}), components of single-site susceptibility tensors, ${\bm \chi}^{(i)}_{\kappa, j}$, are calculated numerically in the temperature range $1 - 50$~K with a step size of 0.25~K.
The average magnetic moments of the Er$^{3+}$ ions are determined according to the following expressions (here, similar to the notation introduced above, $\kappa=$ R1, R2; $j = \rm{I - IV}$, $\langle \bm{m}_{\kappa, \rm{I}} \rangle = \langle \bm{m}_{\kappa, \rm{IV}} \rangle$, $\langle \bm{m}_{\kappa, \rm{II}} \rangle = \langle \bm{m}_{\kappa, \rm{III}} \rangle)$:
\begin{equation}
\langle \bm{m}_{\kappa, j} \rangle = \bm{\chi}_{\kappa, j}^{(1)}\bm{H}_{\kappa, loc1} + \bm{\chi}_{\kappa, j}^{(2)}\bm{H}_{\kappa, loc2} .
\label{eq18_m-kj}
\end{equation}

\begin{table}[tb] 
\caption{Dipolar lattice sums $Q_{\rho \alpha, \rho^\prime \beta}$ (in units of 4$\pi$/3$\nu_c$).}
\begin{center}
\begin{tabular}{|c|c|c|c|c|}
\hline
\hline
$\rho \alpha$, $\rho^\prime \beta$ & $Q_{\rho \alpha, \rho^\prime \beta}$ & $\rho \alpha$, $\rho^\prime \beta$ & $Q_{\rho \alpha, \rho^\prime \beta}$ \\ \hline                                 
$\rho x$, $\rho x$	& -4.116	& $\rho x$, $\rho +4x$	& -1.139    \\
$\rho y$, $\rho y$	& -4.781	& $\rho y$, $\rho +4y$	& 4.144     \\
$\rho z$, $\rho z$	& 11.897	& $\rho z$, 7$z$		& 0.215     \\
1$x$, 2$x$		& 0.629	& 1$x$, 7$y$			& 2.809     \\
1$y$, 2$y$		& 2.370	& 1$x$, 8$x$			& 3.092     \\ 
1$x$, 3$x$		& -1.432	& 1$y$, 8$y$			& -0.084    \\
1$y$, 3$y$		& 3.717	& 1$z$, 8$z$			& -0.008    \\
1$z$, 3$z$		& 0.715	& 1$x$, 8$y$			& 0.006     \\
1$x$, 3$y$		& -4.657	& 5$x$, 6$x$			& 0.634     \\
1$x$, 4$x$		& 2.310	& 5$y$, 6$y$			& 2.366     \\ 
1$y$, 4$y$		& 0.694	& 5$x$, 7$x$			& -1.471    \\
1$x$, 6$x$		& 6.777	& 5$y$, 7$y$			& 3.797     \\
1$y$, 6$y$		& -4.082	& 5$z$, 7$z$			& 0.674     \\
1$z$, 6$z$		& 0.305	& 5$x$, 7$y$			& -4.526    \\
1$x$, 6$y$		& -0.045	& 5$x$, 8$x$			& 2.352     \\
1$x$, 7$x$		& -1.272	& 5$y$, 8$y$			& 0.653     \\
1$y$, 7$y$		& 4.058	& 5$z$, 8$z$			& -0.005    \\ \hline
\end{tabular}
\end{center}
\label{TableIV}
\end{table}   
                              
Such a cluster approach allows us to (at least partly) account for quantum correlations between the magnetic moments. 
Neglecting fluctuations of magnetic moments of the ions outside a fixed cluster (specifically, we consider a cluster containing the Er$^{3+}$ ions at magnetically equivalent R1 sites, belonging to the sublattices with numbers $\rho =1, 3$), we can write the local magnetic fields as follows:
\begin{eqnarray}
\label{eq19_ham-R1l1} 
\bm{H}_{{\rm R1}, loc 1} & = & \bm{H}_{{\rm R1}, 0} - \bm{J}_r^{(ex)}\langle \bm{m}\rangle_{\rm R1} + \Delta \bm{H}_{\rm R1} , \\
\bm{H}_{{\rm R1}, loc 2} & = & \bm{H}_{{\rm R1}, 0} + \bm{J}_r^{(dd)}\langle \bm{m}\rangle_{\rm R1},
\label{eq20_ham-R1l2}
\end{eqnarray}
where 
\begin{eqnarray}
\nonumber
\bm{H}_{{\rm R1}, 0} 	& = & \bm{H} - \bm{H}_D + \sum\limits_{\rho =1}^8 (\bm{Q}_{1 , \rho} - \bm{J}_{1, \rho}^{(ex)}) \langle \bm{m}_\rho \rangle \\*
				& + & (\bm{J}_r^{(dd)}+\bm{J}_c^{(dd)} - \bm{J}_c^{(ex)}) \langle \bm{m}\rangle_{\rm R1}.
\label{eq21_ham-R1-0}
\end{eqnarray}
Here $\bm{H}_{D}$ is the demagnetizing field, and $\langle \bm{m} \rangle_{\kappa}$ are magnetic moments of the Er$^{3+}$ ions at the sites $\kappa$ induced by the external field $\bm{H}$ (a more rigorous definition is given below). 
When $\bm{H} \parallel c$, only the $z$~components of the average single-ion magnetic moments are nonzero, and $\langle m_{\rho, z} \rangle = \langle m \rangle_{{\rm R1}, z}$ for $\rho$ from 1 to 4, $\langle m_{\rho, z} \rangle = \langle m \rangle_{{\rm R2}, z}$ for $\rho$ from 5 to 8. 
For the magnetic field $\bm{H} \parallel a$ and $\bm{H} \parallel b$, the calculations of the local fields are more complicated because the single-ion susceptibility tensors have nonzero off-diagonal components, and the magnetic moments are tilted away from the external magnetic field direction in the $ab$~plane. 
Specifically, for $\bm{H} \parallel a$, the components of the magnetic moments satisfy the relations
$\langle m_{\rho, x} \rangle = \langle m \rangle_{{\rm R1}, x}$ for $\rho$ from 1 to 4,
$\langle m_{\rho, x} \rangle = \langle m \rangle_{{\rm R2}, x}$ for $\rho$ from 5 to 8,
$\langle m_{1, y} \rangle = -\langle m_{2, y} \rangle = \langle m_{3, y} \rangle = -\langle m_{4, y} \rangle = \langle m \rangle_{{\rm R1}, y}$ and
$\langle m_{5, y} \rangle = -\langle m_{6, y} \rangle = \langle m_{7, y} \rangle = -\langle m_{8, y} \rangle = \langle m \rangle _{{\rm R2}, y}$. 
Similar relations, with the rearranged indices $x$ and $y$, are valid for $\bm{H} \parallel b$. 
The matrices $\bm{J}^{(ex)}_{\rho,\rho^\prime}$ in Eq.~(\ref{eq21_ham-R1-0}) correspond to the exchange interactions between the next-nearest-neighbors and between the Er1 and Er2 ions. 
The nonzero values of the dipolar lattice sums
\begin{equation}
Q_{\rho\alpha, \rho^\prime \beta} = \sum\limits_{L(n_1, n_2, n_3)} \frac{(-r_{L\rho \rho^\prime}^2 \delta_{\alpha \beta } + 3x_{L\rho \rho ^\prime \alpha} x_{L\rho \rho^\prime \beta })}{r_{L\rho \rho ^\prime}^5}
\label{eq22_dipolar-lattice}
\end{equation}
(here $x_{L\rho\rho^\prime\alpha}$ is the $\alpha$-component of the vector $\bm{r}_{L\rho\rho^\prime} = n_1\bm{a} + n_2\bm{b} + n_3\bm{c}+\bm{r}_\rho - \bm{r}_{\rho^\prime}$, and $n_1, n_2, n_3$ are integers) are presented in Table~\ref{TableIV} in units of $4\pi/3v_{c}$, where $v_c = abc$ is the unit cell volume).
According to the crystal lattice symmetry $\bm{Q}_{2,3} = \bm{Q}_{1,4};~\bm{Q}_{2,4} = \bm{Q}_{1,3};~\bm{Q}_{3,4} = \bm{Q}_{1,2};~\bm{Q}_{6,7} = \bm{Q}_{5,8};~\bm{Q}_{6,8} = \bm{Q}_{5,7};~\bm{Q}_{7,8} = \bm{Q}_{5,6};~\bm{Q}_{2,5} = \bm{Q}_{3,8} = \bm{Q}_{4,7} = \bm{Q}_{1,6};~\bm{Q}_{2,7} = \bm{Q}_{3,6} = \bm{Q}_{4,5} = \bm{Q}_{1,8};~\bm{Q}_{2,8} = \bm{Q}_{3,5} = \bm{Q}_{4,6} = \bm{Q}_{1,7}$. 
The additional magnetic field $\Delta \bm{H}_{\rm R1}$ introduced in Eq.~(\ref{eq19_ham-R1l1}) is determined from the condition of the magnetic equivalence of all ions in the cluster, namely
\begin{equation}
\langle \bm{m}_{\kappa, \rm{I}} \rangle = \langle \bm{m}_{\kappa, \rm{II}} \rangle = \langle \bm{m}_{\kappa, \rm{III}} \rangle = \langle \bm{m}_{\kappa, \rm{IV}}\rangle .
\label{eq23_mag-equiv}
\end{equation}
This condition renormalizes the single-site susceptibilities, which now take the following form:
\begin{eqnarray}
\nonumber
\bm{\chi}_{\kappa}^{(s)} 	& = & \bm{\chi}_{\kappa, \rm{I}}^{(1)}(\bm{\chi}_{\kappa, \rm{I}}^{(1)} - \bm{\chi}_{\kappa, \rm{II}}^{(1)})^{-1}\bm{\chi}_{\kappa, \rm{II}}^{(2)} \\*
					& - & \bm{\chi}_{\kappa, \rm{II}}^{(1)}(\bm{\chi}_{\kappa, \rm{I}}^{(1)} - \bm{\chi}_{\kappa, \rm{II}}^{(1)})^{-1}\bm{\chi}_{\kappa, \rm{I}}^{(2)} .
\label{eq24_1site-chi}
\end{eqnarray}
The components of the bulk dc susceptibility tensor $\bm{\chi} = N_A (\bm{\chi}_{\rm R1}+\bm{\chi}_{\rm R2})/2$ (per mole of Er, where $N_A$ is the Avogadro constant) are obtained from solutions of a system of self-consistent linear equations which determine the susceptibilities $\bm{\chi}_{\kappa}$ of the Er$^{3+}$ ions at the R1 and R2 sites:
\begin{eqnarray}
\langle \bm{m}\rangle_{\rm R1} & = & \bm{\chi}_{\rm R1}\bm{H} = \bm{\chi}_{\rm R1}^{(s)}\bm{H}_{{\rm R1}, loc2}(\langle \bm{m}\rangle_{\rm R1}, \langle \bm{m} \rangle_{\rm R2}) , \\
\langle \bm{m}\rangle_{\rm R2} & = & \bm{\chi}_{\rm R2}\bm{H} = \bm{\chi}_{\rm R2}^{(s)}\bm{H}_{{\rm R2}, loc2}(\langle \bm{m}\rangle_{\rm R1}, \langle \bm{m} \rangle_{\rm R2}).
\label{eq25_1site-chi-eqs}
\end{eqnarray}

In the temperature range of $1-50$~K, the computed susceptibilities describing the response of \seo\ samples to external magnetic fields directed along the principal crystallographic axes agree well with the experimental data, as shown in Fig.~\ref{Fig8_chi}.
Relatively small differences between the measured and simulated susceptibilities along the $a$ and $b$ axes become noticeable only at temperatures below 3~K. 
It should be noted that the broad maximum observed in the susceptibility along the $a$~axis (at temperatures close to $T_{\rm N} = 0.75$~K) is shifted in the calculated temperature dependence of $\chi_{aa}$ to 1.25~K. 
The results of calculations hint that the low-temperature broad peaks in $\chi_{aa}$ and $\chi_{bb}$ are related to the intrinsic properties of the Er2-ladders (\textit{i.e.}, the geometrically frustrated exchange interactions), and that these peaks could perhaps be observed more clearly if the temperature range was extended to below 0.5~K despite the magnetic ordering of the Er1-ladders.
The parameters of the exchange and dipolar interactions between the Er$^{3+}$ ions in the clusters used in the calculations are presented in Table~\ref{TableV} (note that in our model $\bm{J}_r = \bm{J}_r^\prime$). 
From the data, it can readily be seen that the introduced anisotropic exchange interactions are comparable with the dipolar ones. 
The additional contributions to the local magnetic fields due to the exchange interactions (defined by Eq.~\ref{eq21_ham-R1-0}) are estimated from the fitting procedure.
Namely, they are determined by the parameters
$\sum\limits_{\rho =1,3} J_{1x, \rho x}^{(ex)}=13.4$,
$\sum\limits_{\rho =1,3} J_{1y, \rho y}^{(ex)}=\sum\limits_{\rho =5,7} J_{5y,\rho y}^{(ex)} = \sum\limits_{\rho =5, 7} J_{5z, \rho z}^{(ex)} = 8.9$,
$\sum\limits_{\rho =1,3} J_{1z,\rho z}^{(ex)}=14.5$,
$\sum\limits_{\rho =5,8} J_{1x,\rho x}^{(ex)}=\sum\limits_{\rho =1,4} J_{5x,\rho x}^{(ex)} = 17.8$,
$\sum\limits_{\rho =5,7} J_{5x,\rho x}^{(ex)}=31.2$ (in the units of 10$^{-3}$~cm$^{-1}/\mu_{\mathrm B}^2$).

\begin{table}[tb] 
\caption{Parameters of the exchange ($J_r^{(ex)}$, $J_c^{(ex)}$) and dipole-dipole ($J_r^{(dd)}$, $J_c^{(dd)}$) interactions (in units of 10$^{-3}$ cm$^{-1}$/$\mu_{\mathrm B}^2$) between the Er$^{3+}$ ions in the R1 and R2 ladders.}
\begin{center}
\begin{tabular}{|l|c|c|c|c|}
\hline
\hline
						&  \multicolumn{2}{|c|}{R1}& \multicolumn{2}{|c|}{R2} \\ \hline
$J_{\alpha \beta}^{(\ldots)}$		& $(ex)$	& $(dd)$	& $(ex)$	& $(dd)$		\\ \hline
$J_{r,xx}^{(\ldots)}$				& 2.5		& 4.1		& 28		& 4.2			\\
$J_{r,yy}^{(\ldots)}$				& 2.5		& -7.1	& 7.5		& -7.2		\\
$J_{r,xy}^{(\ldots)}=J_{r,yx}^{(\ldots)}$& 0		& 10.3	& 0		& 10.0		\\
$J_{r,zz}^{(\ldots)}$				& 24.0	& 3.0		& 7.5		& 3.0			\\
$J_{c,xx}^{(\ldots)}=J_{c,yy}^{(\ldots)}$& 5.0	& 11.2	& 5.0		& 11.2		\\
$J_{c,zz}^{(\ldots)}$				& 5.0		& -22.4	& 5.0		& -22.4		\\ \hline			
\end{tabular}
\end{center}
\label{TableV}
\end{table} 

At low temperatures, the energy of interaction determined by the parameter $J_{\rho \alpha, \rho^\prime \beta}^{(ex)}$ is actually a bilinear function of the corresponding $g$-factors of the ground state.
Remembering the large value of $g_{xx}$ of the Er$^{3+}$ ions at the R2 sites, we conclude that the peculiar magnetic properties of these ions (and particularly, the absence of long-range order) are caused by strong \afm\ interactions between the ions forming edge-sharing triangles in the R2 ladders. 
In contrast, in the R1 ladders, a ferromagnetic dipolar interaction along the legs prevails due to the large value of the $g$-factor $g_{zz}$, and induces the long-range magnetic order seen at low temperatures.

The reliability of the parameters presented above and in Table~\ref{TableV} is confirmed by the good agreement between the calculated and measured field dependencies of the magnetization of \seo\ for the magnetic fields directed along different crystallographic axes as shown in Fig.~\ref{Fig9_M_Er}. 
Calculations of the average self-consistent magnetic moments at 5~K were performed using density matrices corresponding to the Hamiltonians (\ref{eq16_ham-c}) of the four-particle clusters in the R1 and R2 ladders and the procedure of subsequent approximations. 
The results of calculations (in Fig.~\ref{Fig9_M_Er}) clearly demonstrate that there is a strong magnetic anisotropy of the easy-axis type along the $c$~axis at the R1 sites and along the $a$~axis at the R2 sites.

\section{Summary}
To summarize, we report the results of a systematic study of the behavior of the magnetic Er$^{3+}$ ions in \seo\ as well as in a lightly doped nonmagnetic analogue, \syoE.  
The observed EPR spectra in the single crystals of \syoEhalf\ established very anisotropic $g$-factors for the Er$^{3+}$ ions in the two crystallographically inequivalent sites.
The CF energies of the Er$^{3+}$ ions substituting for the Y$^{3+}$ ions were determined by making use the methods of the site-selective laser spectroscopy.
This combination of the EPR and spectral studies was further complemented by inelastic neutron scattering as well as the measurements of the field dependencies of magnetization in the paramagnetic phase of the concentrated system, \seo.
The obtained data allowed us to develop a theoretical model to describe the electronic structure of the Er$^{3+}$ ions both in the magnetically dilute and concentrated samples.
Although operating with a large number of independent parameters it was not possible to determine all of their values unambiguously, we are able to estimate most of these parameters from the literature data or from simulations based on known models developed for CF theory.
The sets of CF parameters obtained for the Er$^{3+}$ ions at the two crystallographically inequivalent positions are related to the crystallographic axes, contrary to the results of the CF simulations for the impurity Eu$^{3+}$ ions in \sRo\ ($R=$~Y, Gd, In) reported in Ref.~\cite{Taibi_1993}.

It is rather important to be able to separate the single-ion effects from the collective interactions for the overall understanding of the complex low-temperature magnetic properties of the concentrated strontium rare earth oxides \sRo\ (as well as the related family of barium rare earth oxides~\cite{Doi_2006,Besara_2014,Aczel_2014}).
We hope that the approach adopted in this paper and that has proven successful for \seo, could be applied to other compounds.
In particular, our initial simulations show an Ising type magnetic anisotropy along the $c$~axis and close to the $b$~axis in the ground quasidoublet states of the Ho$^{3+}$ ions at the R1 and R2 sites, respectively, in \sho.
A more detailed analysis of the spectral and magnetic properties of the Ho$^{3+}$ ions in \syo\ and \sho\ is in progress.

\section*{ACKNOWLEDGMENTS}
The work of R.V.~Yusupov, D.G.~Zverev, R.~Batulin, I.F.~Gilmutdinov and A.G.~Kiiamov was funded by the subsidy of the Ministry of Education and Science of the Russian Federation allocated to Kazan Federal University.
B.Z.~Malkin is grateful for support from the Russian Foundation for Basic Research (grant No. 14-02-00826).
\bibliography{SrLn2O4_all}
\end{document}